\pdfoutput=1
\documentclass[journal=jancac3,manuscript=article]{achemso}

\usepackage[sort&compress,numbers,super]{natbib} 
\usepackage{achemso} 
\usepackage{natbib}

\usepackage[version=3]{mhchem} 
\usepackage{gensymb}

\author{Mihail I. Petrov}
\affiliation{The International Research Centre for Nanophotonics and Metamaterials, ITMO University,  Birzhevaja line 14, 199034 St. Petersburg, Russia}
\author{Sergey V. Sukhov}
\affiliation{CREOL, The College of Optics and Photonics, University of Central Florida, 4000 Central Florida Blvd., 32816 Orlando, Florida, USA}

\alsoaffiliation{The International Research Centre for Nanophotonics and Metamaterials, ITMO University,  Birzhevaja line 14, 199034 St. Petersburg, Russia}
\alsoaffiliation{Kotel'nikov Institute of Radio Engineering and Electronics of Russian Academy of Sciences (Ulyanovsk branch), Spasskaya str. 14, 432011 Ulyanovsk,  Russia}

\author{Andrey A. Bogdanov}

\affiliation{The International Research Centre for Nanophotonics and Metamaterials, ITMO University,  Birzhevaja line 14, 199034 St. Petersburg, Russia}

\alsoaffiliation{Ioffe Institute, Politekhnicheskaya str. 26, 194021 St. Petersburg,   Russia}

\alsoaffiliation{Peter the Great St.Petersburg Polytechnic University, Politekhnicheskaya str. 29, 195251 St. Petersburg, Russia}

\author{Alexander~S.~Shalin}

\affiliation{The International Research Centre for Nanophotonics and Metamaterials, ITMO University,  Birzhevaja line 14, 199034 St. Petersburg, Russia}
\alsoaffiliation{Kotel'nikov Institute of Radio Engineering and Electronics of Russian Academy of Sciences (Ulyanovsk branch), Spasskaya str. 14, 432011 Ulyanovsk,  Russia}

\alsoaffiliation{Ulyanovsk State University, Lev Tolstoy str. 42,  432017 Ulyanovsk, Russia}

\email{alexandesh@gmail.com}

\author{Aristide~Dogariu}
\affiliation{CREOL, The College of Optics and Photonics, University of Central Florida, 4000 Central Florida Blvd., 32816 Orlando, Florida, USA}

\title{Surface plasmon polariton assisted \\ optical pulling force\footnote{This is the pre-peer reviewed version of the following article: [M.I.~Petrov, S.V.~Sukhov, A.A.~Bogdanov, A.S.~Shalin, A.~Dogariu, Laser \& Photonics Reviews. DOI:10.1002/lpor201500173 (2015)], which has been published in final form at http: http://onlinelibrary.wiley.com/doi/10.1002/lpor.201500173/abstract. This article may be used for non-commercial purposes in accordance with Wiley Terms and Conditions for Self-Archiving.This article may be used for non-commercial purposes in accordance with Wiley Terms and Conditions for Self-Archiving.}}

\abbreviations{SPP}
\keywords{Optical manipulation, optical forces, optical tractor beams, surface plasmon polaritons}

\begin{document}

\begin{abstract}
  We demonstrate both analytically and numerically the existence of optical pulling forces acting on particles located near plasmonic interfaces. Two main factors contribute to the appearance of this negative reaction force. The interference between the incident and reflected waves induces a rotating dipole with an asymmetric scattering pattern while the directional excitation of surface plasmon polaritons (SPP) enhances the linear momentum of scattered light. The strongly asymmetric SPP excitation is determined by spin-orbit coupling of the rotating dipole and surface plasmon polariton. As a result of the total momentum conservation, the force acting on the particle points in a direction opposite to the incident wave propagation. We derive analytical expressions for the force acting on a dipolar particles placed in the proximity of plasmonic surfaces. Analytical expressions for this pulling force are derived within the dipole approximation and are in excellent agreement with results of electromagnetic numerical calculations. The forces acting on larger particles are analyzed numerically, beyond the dipole approximation.
\end{abstract}

\section{Introduction}
In spite of wide popularity of volume optical manipulation (mainly, for biological research),\cite{grier2003revolution} it has major drawbacks if applied to lab-on-a-chip platforms and micro-fluidic devices. On the other hand, manipulation with evanescent fields and surface waves is compatible with co-planar integrations needed for miniaturization of optical devices.\cite{reece2006near} The high spatial localization of evanescent field might enable to extend optical manipulation down to nanometer scale.\cite{nieto2004near} The optical forces acting in a system can be significantly amplified if Surface Plasmon Polaritons (SPP) on metal-dielectric interfaces are used.\cite{volpe2006surface,quidant2008surface} In addition to enhancing the magnitude of optical forces, SPP can achieve further spatial confinement down to nanometer scale.\cite{volpe2006surface,xu2002surface}

Various aspects of SPP manipulation were already considered. Plasmonic amplification was used for trapping or acceleration of nano-sized particles using high gradients of local fields, \cite{volpe2006surface, novotny1997theory, righini2007parallel, shalin2013plasmonic}  and for modulation plasmonic signal by oscillating nanoparticle in V-shaped waveguide \cite{Shalin2014}. In other situations, the propagating SPP waves were used for transportation of particles using localized scattering forces.\cite{wang2009propulsion} Combination of gradient and scattering forces allows to construct robust SPP trap for mesoscopic metallic particles.\cite{min2013focused} Normal to the surface force was recently considered also in flat plasmonic and metamaterial structures. \cite{Zhang2012,Liu2011} Here we discuss the unusual situation when nonconservative force from SPP wave creates attractive force on small particles acting along the interface towards the source of light (plasmonic ''tractor beam'').

Recently, nonconservative pulling forces (''tractor beams'') attracted considerable attention because of an unusual behavior: they act in a direction opposite to the optical wave propagation, toward the source of light. Many types of tractor beams have already been suggested\cite{dogariu2013optically} most of them being based on either structuring the incident field\cite{Chen2015,Sukhov2011,Lee2010,novitsky2011single,Novitsky2014,chen2011optical} or on modifying the surroundings of the manipulated object.\cite{Shalin2015optical,salandrino2011reverse,Bogdanov2015a} For instance, backward surface waves existing in certain structures can provide means for such manipulation.\cite{maslov2014resonant} In other schemes, the negative optical forces appear because of the enhancement of optical linear momentum after interaction of light with an object. Attraction force can appear, for instance, when a particle is placed in the vicinity of a waveguide because of mode transformation inside the waveguide.\cite{intaraprasonk2013optical}.

Another remarkable illustration of optical attraction is the case of micrometer-size particles floating on a liquid surface and driven towards the source of light due to the natural enhancement of light momentum in optically denser media.\cite{kajorndejnukul2013linear,Qiu2015} Of course, in terms of application in nano-photonic devices, this concept would be more interesting if it could be extended to solid substrates. The direct application of this method to solid surfaces is impossible because the manipulated objects reside entirely outside the substrate. However, such an object can still interact with the solid substrate by converting some of its evanescent field into waves that propagate into the substrate. This interaction effectively changes the scattering pattern of the object redistributing the nonconservative reaction forces. If the photons' momentum in substrate increases, the object should experience a backward force to compensate this surge of momentum. For metallic surfaces, this increase of momentum can occur as a result of SPP excitation, which have large wavenumbers and, therefore, large linear momenta. In this paper we explore this idea in detail and provide both analytical and numerical results confirming the possibility of creating attractive optical forces for particles placed on solid surfaces. Particular attention is devoted to plasmonic interfaces.

\section{Results and discussion}

\subsection{Forces acting on a point dipole near interface}

\begin{figure}
  \centering
   \includegraphics[scale=0.5]{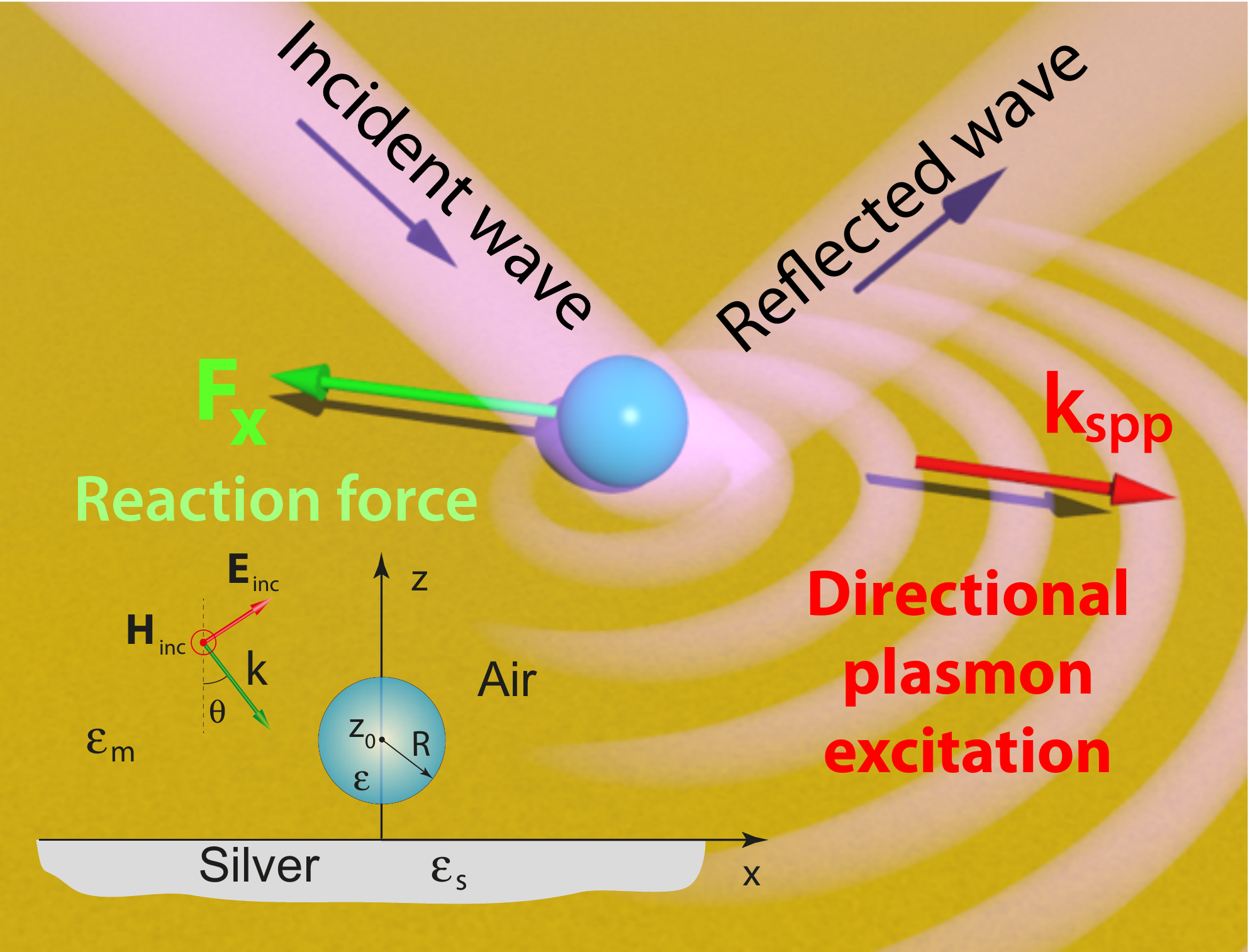}
  \caption{Directional excitation of surface plasmon polaritons by a small particle placed in the vicinity of a metal surface. The recoil force created by plasmon polaritons on the particle can be directed opposite to the propagation of incident light. The insert depicts the geometry of the problem.}
  \label{fig_Fig1}
\end{figure}

We start by examining a simplified situation for which an analytical solution is possible: the case of a dipole located in the vicinity of a solid substrate. We consider a spherical dielectric nanoparticle with permittivity $\varepsilon$  and radius $R$   located inside medium with permittivity $\varepsilon_\text{m}$. The particle is placed at the height  $z_\text{0}$  above a substrate with permittivity $\varepsilon_\text{s}$, as shown in~\ref{fig_Fig1}. The incident wave with wavelength $\lambda$   impinges onto the surface at the angle~$\theta$. Assuming that the nanoparticle is small in comparison to the wavelength, $R\ll\lambda$, one can consider it as a point electric dipole with polarizability $\alpha_0$.

The general expression for the optical force acting on a point dipole can be written as\cite{ashkin1983stability,chaumet2000time}   
\begin{equation}
\mathbf{F}=\frac{1}{2}\sum_{i}\text{Re}\left(p_{i}^*\nabla E_{i}^\text{loc}\right),
\end{equation}
where $\mathbf{p}=\alpha_0\mathbf{E}^\text{loc}$  is the induced dipole moment, and $\mathbf{E}^\text{loc}$ is the local electrical field. The local electrical field above the substrate 
\begin{equation}
\mathbf{E}^\text{loc}(\mathbf{r})=\mathbf{E}^0(\mathbf{r})+\mathbf{E}^\text{D}(\mathbf{r})
\label{eq2}
\end{equation}	 	
is the superposition of two components: the field $\mathbf{E}^0$, which includes both the incident and reflected waves, and the scattered field $\mathbf{E}^\text{D}$, which includes field induced by the point dipole, and local fields generated in the substrate (see~\ref{fig_Fig1}). The second component can be found to be $\mathbf{E}^\text{D}=\omega^2\mu_0\hat G^\text{R} (\mathbf{r},\mathbf{r}_0)\mathbf{p}$  in terms of the dyadic Green's function $\hat G^\text{R}$  of the dipole-interface system\cite{novotny2012principles}  (see Supporting Information A for explicit expressions). Thus, the total induced dipole moment  $\mathbf{p}=\hat\alpha\mathbf{E}^0$ can be written using an effective polarizability tensor  $\hat\alpha$ which also includes the nanoparticle interaction with the substrate. This effective polarizability is defined as
\begin{equation}
\hat\alpha=\alpha_0\left[\mathbf{I}-\alpha_0\omega^2\mu_0\hat G^\text{R}(\mathbf{r}_0,\mathbf{r}_0)\right]^{-1},
\label{Eq:eq3}
\end{equation}			 	
where $\mathbf{I}$ is the unity tensor, $\mu_{0}$ is the magnetic permittivity of vacuum, and $\omega$ is the frequency of incident wave. The diagonal elements have the following form:  
\begin{equation}
\alpha_\text{xx}=\alpha_\text{yy}=\frac{\alpha_0}{1-\alpha_0\omega^2\mu_0  G_\text{xx}^\text{R}(\mathbf{r}_0,\mathbf{r}_0)}, \ \ \alpha_\text{zz}=\frac{\alpha_0}{1-\alpha_0\omega^2\mu_0 G_\text{zz}^\text{R}(\mathbf{r}_0,\mathbf{r}_0)}.
\label{Eq:eq4}
\end{equation}
As a result of the field superposition in \ref{eq2}, the optical force acting along interface could also be separated into two terms $F_\text{x}=F_\text{x}^0+F_\text{x}^\text{D}$ where
\begin{equation}
F_\text{x}^{0(D)}=\frac{1}{2}\text{Re}\left(\mathbf p^*\frac{\partial\mathbf E^{0(D)}}{\partial x}\right). 
\label{Eq:eq5}
\end{equation}
Term $F_{x}^{0}$ corresponds to the x-component of the total radiation pressure generated by the incident and the reflected waves while $F_\text{x}^\text{D}$ corresponds to the interaction of the dipole with its own retarded field. This separation is just a matter of convenience because, in fact, $F_\text{x}^{0}$ also accounts for the interaction with the substrate through the polarizability tensor in \ref{Eq:eq4}. It is important to note that, as long as the system depicted in~\ref{fig_Fig1} remains invariant with respect to the motion of the dipole along its surface,  $F_\text{x}$ represents a purely nonconservative force. After some algebra, the expression for the $F_{x}^{0}$  becomes  
\begin{equation}
F_\text{x}^{0}=\frac{1}{2}k_x\left[\text{Im}\left(\alpha_\text{xx}\right)\left|E_\text{x}^0\right|^2+\text{Im}\left(\alpha_\text{zz}\right)\left|E_\text{z}^0\right|^2\right]
\label{Eqeq6}
\end{equation}
with $k_x=k\sin\theta$,   and $k=\varepsilon_{\text{m}}^{1/2}\omega/c$   being the wavenumber in the upper half-space. The right hand side in \ref{Eqeq6} is always positive because  $\text{Im}\left(\alpha_{xx,zz}\right)>0$ as required by energy conservation for $\exp(-i\omega t)$ time dependence. Thus, the force component $F_\text{x}^{0}$ acts always along the direction of incident wave propagation. 

It can be shown that the other force component,  $F_\text{x}^{D}$, corresponding to the dipole interaction with its retarded field (see Supporting Information B for details) can be written as  
\begin{equation}
F_\text{x}^\text{D}=-\frac{k^2}{\varepsilon_0}\text{Im}\left(p_\text{x}^*p_\text{z}\right)\text{Im}\left(\partial_x G^\text{R}_{xz}\right),
\label{Eq:eq7}
\end{equation}
where $\varepsilon_{0}$ is the dielectric  permittivity of vacuum. 

This formula is one of the main results of our paper. It indicates that, interestingly, this force component is nonzero when there is a phase difference between the x- and z-components of the dipole moment. Depending on this phase difference, the force $F_\text{x}^\text{D}$  can be either positive or negative, which is of particular interest for our discussion. We also note that, because  $F_\text{x}^\text{D}$ depends only on the x- and z-components of the dipole moment, it is sufficient to analyze only the case of $p$-polarized incident waves. Remarkably, although $F_\text{x}^\text{D}$  acts on a dipolar particle located close to an interface, this force cannot be predicted by the electrostatic theory of particle-surface interaction. This is due to the fact that $F_\text{x}^\text{D}$  is nonconservative and its description requires taking into account phases of the field which are neglected in the electrostatic approximation.

The phase difference between  two components of  dipole moment   (leading to dipole's rotation) is a key requirement for the asymmetry of the dipole's scattering pattern for particles located in the vicinity of interfaces. It also leads to the directional excitation of waves in a substrate due to the spin-orbit coupling effect. The situation is typical for objects placed close to dielectric media,\cite{haefner2009spin,petersen2014chiral} metal surfaces,\cite{xi2013controllable,mueller2013asymmetric,rodriguez2013near} or metasurfaces.\cite{Yermakov2015} This asymmetry produces a linear momentum imbalance that should be compensated by additional force  $F_\text{x}^\text{D}$ acting on the scattering dipole. When the dipole is located very close ($kz_0\ll1$)  to non-absorbing dielectric media with real reflection coefficient, the $p_\text{x}$  and $p_\text{z}$  oscillate synchronously and, consequently,  $F_\text{x}^\text{D}=0$. However, for metallic interfaces, the interference of incident and reflected waves can induce a phase shift of almost $\pi/2$ between the phases of the $p_\text{x}$  and $p_\text{z}$  generating, therefore, a rotating electric dipole. When the scattering from this rotating dipole excites SPP wave propagating predominantly in forward direction (along x-axis), one can anticipate an increase of the momentum of photons in the substrate.  Then, due to linear momentum conservation law, the object should experience negative force $F_\text{x}$  to compensate this momentum increase. 

The analysis of $\text{Im}\left(p_\text{x}^*p_\text{z}\right)$  allows us to understand the dependence of $F_\text{x}^\text{D}$  on the angle of incidence. For example, for dipole's locations close to the interface ($kz_0\ll1$) one gets (see Supporting Information B)
\begin{equation}
\text{Im}\left(p_\text{x}^*p_\text{z}\right)=|\alpha_0|^2|\mathbf{E}^0|^2 \sin(2\theta)\text{Im}\left[r_\text{p}(\theta)\right],
\label{Eq:eq8}
\end{equation}
where $r_{p}$ is the Fresnel coefficient for p-polarized wave. It follows from this formula that, in particular,  $F_\text{x}^\text{D}=0$ for both normal and grazing angles of incidence. 
									  
As can be seen in \ref{Eq:eq7}, the magnitude and the direction of the force   depend also on the derivative of the Green's function, which can be expressed through the integral in Fourier space (see Supporting Information B):
\begin{equation}
\text{Im}\left(\partial_x\hat G^\text{R}_{xz}\right)=\frac{1}{8\pi k^2}\text{Im}\left(\int\limits_0^\infty k_\|^3r_\text{p}(k_\|)\exp(2ik_\text{z}z_0)\text{d}k_\|\right),
\label{Eq:eq9}
\end{equation}
where  $k_\|$ and $k_\text{z}$   are the wavevector components lateral and normal to the interface.  The integration is performed over the lateral component of the wave vector in the upper half-space. We note that the integration range $0<k_\|<k$  in \ref{Eq:eq9} corresponds to the contribution from the propagating waves reflected from the surface, whereas the integration over  $k_\|>k$  relates to the contribution from the modes evanescent in the upper space. This integral is evaluated for special cases in the Supporting Information C.
										  
It is well-known\cite{novotny2012principles} that if the condition $\varepsilon_\text{m}+\varepsilon_\text{s}<0$ is fulfilled, an evanescent surface plasmon wave can propagate along the interface. In the integral in \ref{Eq:eq9} it relates to a pole in the reflection coefficient at the SPP wavevector  $k_\text{SPP}=k\left[\varepsilon_\text{m}\varepsilon_\text{s}/(\varepsilon_\text{m}+\varepsilon_\text{s})\right]^{1/2}$. Even though the reflection coefficient is purely real for  $k_\|>k$ , the imaginary part of the integral is nonzero due to the regularization of the integrand divergence (see Supporting Information C). Besides, the Green's function tensor components expressed in terms of similar integrals are incorporated into  $\alpha_\text{xx}$ and combined all-together define the force $F_\text{x}^\text{D}$. When calculated near the surface plasmon (SP) resonance $\text{Re}(\varepsilon_\text{m}+\varepsilon_\text{s})\approx0$, in the case of low-absorbing substrate with $\text{Re}(\varepsilon_{\text{s}})<0$, and far from configurational resonance, this force becomes (see Supporting Information C):
\begin{equation}
F_\text{x}^\text{D}\approx-\frac{k_\text{SPP}^4}{8\varepsilon_0}\frac{k_\text{SPP}^3}{k^3}|\alpha_0|^2|\mathbf{E}^0|^2
\sin(2\theta)\sin\left[2(k_\text{z}z_0+\phi)\right]\exp\left(-2\sqrt{k_\text{SPP}^2-k^2}z_0\right),
\label{Eq:eq10}
\end{equation}
where $\phi$  is related to the phase of reflection coefficient, which in the case of non-absorbing media has the form  $r_\text{p}\approx\exp(2i\phi)$. One can see from \ref{Eq:eq10} that $F_\text{x}^\text{D}$ depends strongly on the SPP wavevector $k_\text{SPP}$ $\left( \propto k_\text{SPP}^7 \exp\left(-2\sqrt{k_\text{SPP}^2-k^2}z_0\right) \right)$, and right at the SP resonance, $k_{SPP}\rightarrow \infty$, the exponential factor zeroes the lateral force for fixed distance $z_{0}$. Thus, the maximal attractive force is spectrally slightly red-shifted with respect to SP resonance. The same expression gives us the understanding of lateral force dependence on the elevation of the particle  above the surface ($z$-coordinate).  Because of the  exponential factor the plasmonic force  exponentially decays with height, and in point dipole approximation goes to zero exactly at the interface, assuming that $\phi\approx \pi$ in the metallic regime of the substrate. Generally, there exists an optimal height, at which  the SPP force reaches its maximal negative value for fixed wavelength and incidence angle. It can be derived from the  \ref{Eq:eq10} and for considered assumption near the SP resonance is described by simple expression $z_{max}\approx 1/\left(2\sqrt{k_\text{SPP}^2-k^2}\right)$, which is a half of SPP penetration depth in the upper medium.

\subsection{SPP enhanced optical pulling forces} 

\begin{figure}
   \centering
   \includegraphics{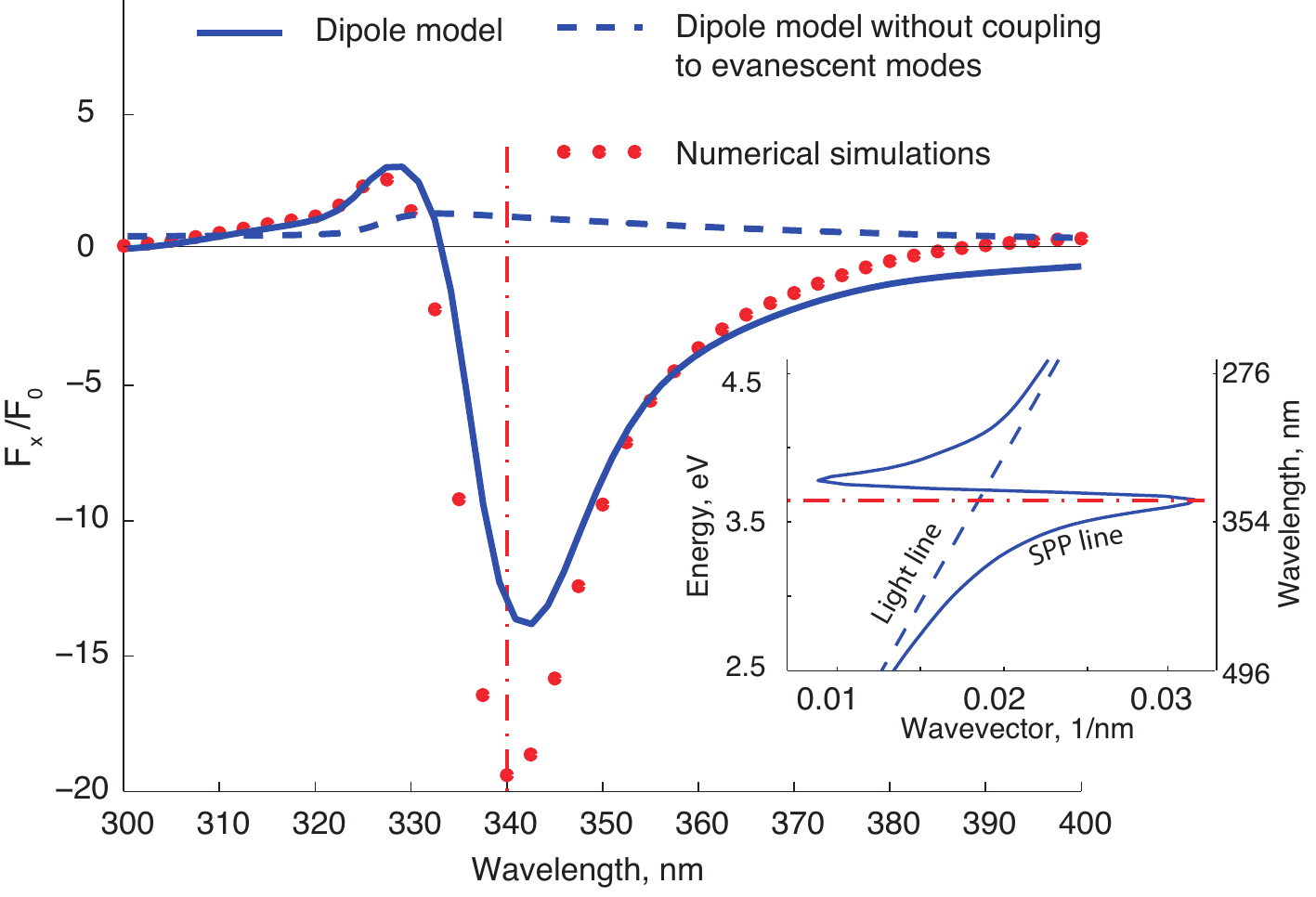} 
   \caption{The spectral dependence of the longitudinal force $F_\text{x}$ acting on a dielectric particle with radius $R=15$~nm and dielectric permittivity $\varepsilon=3$ placed in air above a silver surface at $z_0=27$~nm. Dielectric permittivity of silver is taken from work by~\citeauthor{Johnson1972}\cite{Johnson1972} and the angle of incidence is  $\theta=35\degree$. The solid line corresponds to the dipole model; dotted line corresponds to the numerical calculations. The dashed line corresponds to dipole model  with excluded evanescent modes contribution by integrating the Green's function in the range $0<k_\|<k$ only. Inset: the dispersion curve of SPP at the silver/air interface. The dash-dot line corresponds to SP resonance wavelength.}
   \label{fig_Fig2}
\end{figure}

In this section we find the specific parameters corresponding to overall pulling force and compare analytical predictions to exact numerical calculations. First, as an example, we compute the longitudinal component of the optical force $F_\text{x}$  for a dielectric nanoparticle with permittivity $\varepsilon=3$  (e.g., melamine) and radius $R=15$~nm in air ($\varepsilon_{m}=1$) placed in the proximity of a silver substrate. The force on the dipole is evaluated within the model outlined before and the results are summarized in~\ref{fig_Fig2}. We have normalized the force $F_\text{x}$ with the radiation pressure force $F_0=1/2k|\mathbf{E}^0|^2\text{Im}(\alpha_0)$  acting on the spherical nanoparticle in the absence of a substrate. This force is nonzero as $\text{Im}(\alpha_0)>0$ due to scattering of light, and it is proportional to nanoparticle volume. As can be seen, in the spectral region 330-400 nm, the force $F_\text{x}$  resonantly becomes negative exceeding the radiation pressure force by an order of magnitude. According to the dispersion curve shown in the inset, the localized SPP exists at wavelengths longer than 340 nm (dash-dot line), which fulfills the condition $\text{Re}(\varepsilon_\text{s}+\varepsilon_\text{m})=0$  and explains the observed resonant behavior. Below 340 nm the SPP is not localized and gives contribution to the far-field modes. 
		
To test the analytical model and to further extend this force concept to larger particles, we have evaluated the problem numerically with Comsol Multiphysics (see Methods section for details). The results are shown in~\ref{fig_Fig2} with dotted line. In general, there is good correspondence between the numerical and analytical results; the minor discrepancy relates to the finite size of nanoparticle used in the numerical simulations. Indeed, in the simulations the particle is subjected to fields that are strongly inhomogeneous in the z-direction. As a result,  the location of the induced dipole moment does not coincide with the geometrical centre of the particle and it is effectively located closer to the surface. In fact, the small discrepancy between analytical and numerical results vanishes almost completely if one computes the force for $z_0=26$~nm, which is 1 nm smaller than $z_0$ in~\ref{fig_Fig2}.

	\begin{figure}
   \centering
   \includegraphics{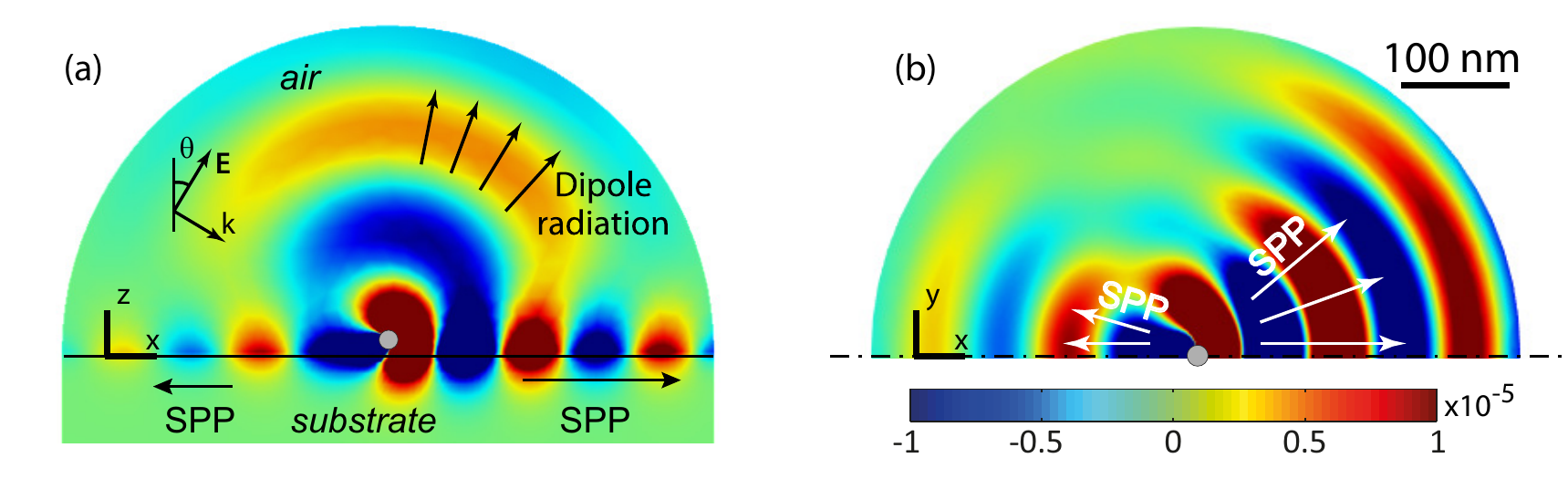} 
   \caption{The distribution of the scattered  $H_z^\text{D}$ field component: (a) the ''side'' view in the x-z plane; (b) the ''top'' view in the x-y plane (only half plane is shown due to symmetry). The results are computed for wavelength 350~nm, and the other parameters of calculation are the same as in~\ref{fig_Fig2}. The field amplitude is normalized over the amplitude of the incident field.}
   \label{fig_Fig3}
\end{figure}

A direct numerical confirmation of the directional SPP excitation is presented in~\ref{fig_Fig3} where the modal SPP structure is evident. Dynamics of the scattered field is available in the Supporting Information E. As discussed earlier, the condition for the directional SPP excitation is the $\pi/2$ phase shift between x- and z-components of dipole moment. This phase difference rapidly grows at the SP resonance at 340~nm, and reaches $\pi/2$ at longer wavelengths (see additional discussion in Supporting Information D), that corresponds to the case of ideally metallic surface as permittivity increases. However, the absence of negative force for longer wavelengths tells us that $\pi/2$ phase difference is not sufficient when conditions for SPP excitation are not met.

\begin{figure}
   \centering
   \includegraphics[scale=0.8]{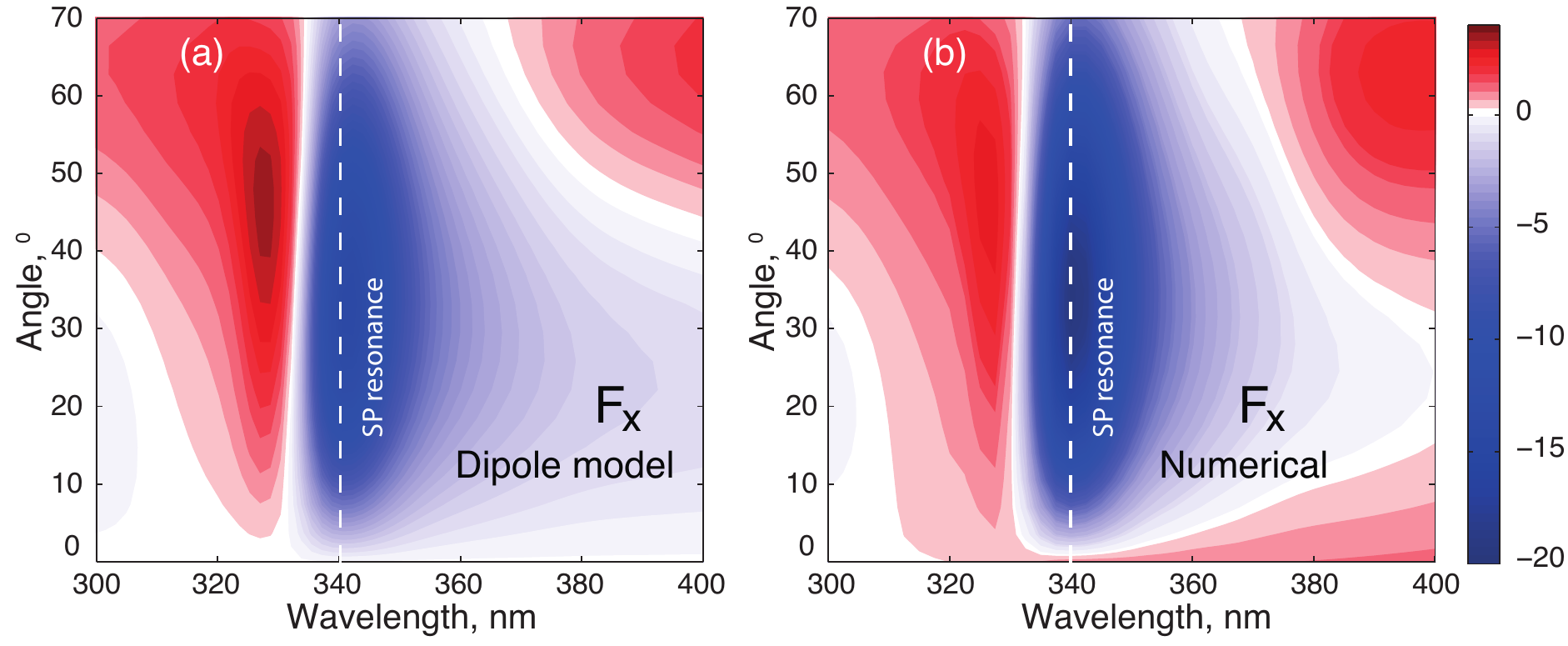} 
   \caption{The x-component of the optical force computed for different wavelengths and incidence angles within the analytical dipole model (a) and numerically (b). The parameters of calculation are the same as in~\ref{fig_Fig2}. The amplitude of the force is given in units of $F_{0}$.}
   \label{fig_Fig4}
\end{figure}

As shown in~\ref{fig_Fig4}a, the force $F_\text{x}$  is negative for a wide range of incidence angles  $\theta$  starting from almost normal incidence and till 70\degree ; it reaches maximum (in absolute value) around 35\degree~incidence. In the numerical results that account for finite size effects, a similar behavior is evident over the wide spectral and angular regions (see~\ref{fig_Fig4}b). A good correspondence between the numerical and analytical results is also observed for the z-component of the optical force (see Supporting Information E).

\begin{figure}
   \centering
   \includegraphics[width=0.9\linewidth]{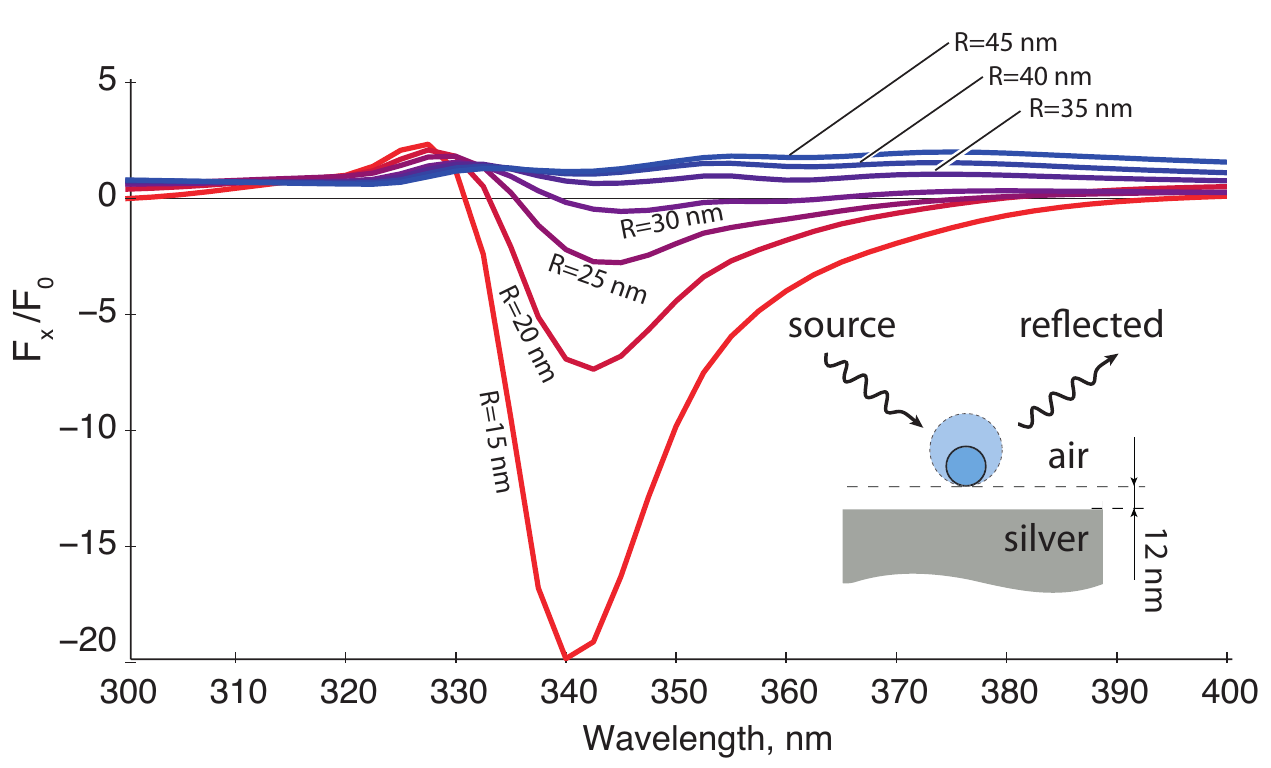} 
   \caption{The x-component of the optical force for nanoparticle radii ranging from 15~nm to 45~nm. Insets: the simulation setting with a constant gap between the nanoparticle and the surface.}
   \label{fig_Fig5}
\end{figure}

\begin{figure}[h]
   \centering
   \includegraphics{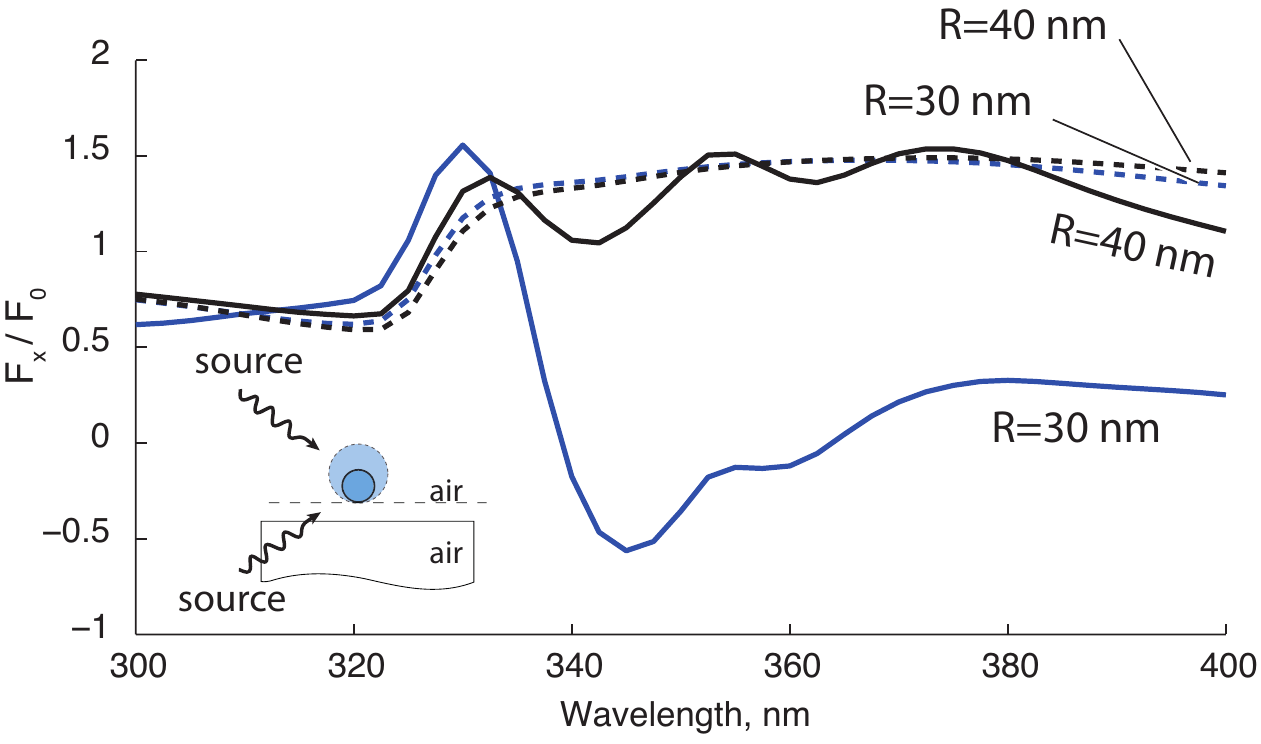} 
   \caption{The $x$-component of the optical force for nanoparticles with radii of 30~nm and~40 nm (solid lines) together with the $x$-component of the force generated only by the incident and reflected waves (dashed lines) without accounting for particle-surface interaction.  Inset: the substrate interaction exclusion by using an additional source that which accounts for the reflected wave. }
   \label{fig_Fig6}
\end{figure}	

Next, we have analyzed the dependence of the x- component of the optical force on the nanoparticle size. The force acting on the nanoparticle was evaluated numerically and its spectral dependence is shown in~\ref{fig_Fig5} for different particle sizes. The results correspond to changing the radius   while keeping constant the 12~nm gap between the nanoparticle and surface as shown in~\ref{fig_Fig5} inset. One can see that for larger particles the lateral force becomes positive, which is to be expected because when increasing the nanoparticle radius we also increase the distance between the surface and the effective dipole placed at the nanoparticle center. As a result, dipole-to-substrate coupling diminishes significantly [see \ref{Eq:eq10}].

We would like to emphasize that the origin of the attractive force discussed in this paper is conceptually new and it should be distinguished from the one discussed in Refs.~\citenum{chen2011optical},~\citenum{novitsky2011single}. Although the experimental verification of that force\cite{brzobohaty2013experimental}  included the reflecting surface, the interaction of the object with this surface did not play any role. The nature of the pulling force in Refs.~~\citenum{chen2011optical},~\citenum{novitsky2011single} is based on the redistribution of scattering due to interaction of multipoles inside the scattering object. Thus, this force would be absent for particles having only electric dipole moment which are the type of particles described in our paper. To assess the influence of the force of Refs.~\citenum{chen2011optical},~\citenum{novitsky2011single}, we performed additional calculations, in which we eliminated the interaction with the substrate using the following simulation procedure: we have excluded the substrate and added second light source. Choosing the amplitude and phase of this second source in a proper way, we fully simulate the wave reflected from the metal surface. The results of this calculation are shown in~\ref{fig_Fig6} for nanoparticles of two sizes. One can clearly see that for a 30~nm nanoparticle interacting with the substrate the force can be negative (solid blue line) but it becomes positive after canceling the influence of the substrate (blue dashed line).  This is a direct evidence that the pulling force proposed in our paper can not be explained by interplay of multipolar momenta as in Refs.~\citenum{chen2011optical},~\citenum{novitsky2011single}. For the nanoparticle of 40~nm radius the interaction with substrates is weaker and the total force (black solid line) almost coincides with the force obtained after substrate exclusion (black dashed line). Some additional oscillations are still present but increasing the nanoparticle size even further results in a smoother variation of the total force. The oscillatory behavior of the force acting on the 40~nm nanoparticle relates to the oscillations of the spatial Fourier harmonics that leads to fluctuating interaction with the substrate.

\section{Conclusions}

In this paper, we presented a theory for the optical force acting on dipolar particles located in the vicinity of a substrate supporting surface waves. The part of the force responsible for the interaction with the substrate depends on a phase between dipole components and allows negative values. That uncovers a new mechanism for creating optical attractive forces (surface assisted tractor beams). In the particular case of metallic surfaces, we have shown that the interference of incident and reflected waves induces an effective rotating dipole. When interacting with the substrate, this rotating dipole produces an asymmetric distribution of the scattered field, which, in turn, leads to a scattering nonconservative force acting in a direction opposite to the incident beam propagation. To overcome radiation pressure from the incident and reflected from a substrate beams, additional directional excitation of surface plasmon polaritons is necessary. The large SPP momentum provides a significant recoil force on the particle and generates a negative force that overcomes the radiation pressure by an order of magnitude. Analytical expressions were derived for this negative surface-assisted force. The theoretical ideas were confirmed by exact numerical calculations. Numerical calculations show that negative forces exist for particles with dimensions up to 60 nm in diameter. This maximum dimension is defined by the extension of the evanescent SPP modes above the surface.

The proposed new concept of surface assisted optical manipulation can be used in integrated photonic circuits, optofluidic devices, and to improve existing methods of optical transport.

\begin{acknowledgement}
This work has been supported in part by Government of the Russian Federation (No. 074-U01), Russian Fund for Basic Research within the Project No. 15-02-01344. The investigation of optical forces distributions has been supported by the Russian Science Foundation (No. 14-12-01227). A.A.B. thanks RFBR (Project No. 14-02-01223), the Federal
Programme on Support of Leading Scientific Schools (NSh-5062.2014.2), and the program of Fundamental Research in Nanotechnology and Nanomaterials of the Russian Academy of Science.
\end{acknowledgement}

\subsection{Supplementary information}
(A) Expressions for the dyadic Green's function that are necessary for calculation of polarizability for a dipole near the surface. (B) Derivation of forces acting on a dipole near the surface. (C) Calculation of the integral entering the expression of derivative of the Green's function with Figure S1. (D) Phase of a dipole moment near the interface with Figure S2. (E) Mapping the angular and spectral dependence of the component of the force normal to the interface with Figure S3. (F) Description of the movie animating SPP dynamics 

\bibliography{references}

\end{document}